\documentclass[twocolumn,showpacs,preprintnumbers,amsmath,amssymb,APSl,prd,nofootinbib,superscriptaddress]{revtex4-1}
\usepackage{dcolumn}
\usepackage{bm}
\usepackage{ifpdf}
\usepackage{hyperref}
\usepackage{dcolumn}
\usepackage{bm}
\usepackage[spanish,english]{babel}
\usepackage{amsfonts}
\usepackage{amssymb}
\usepackage{graphicx}
\usepackage{color}
\usepackage[latin1]{inputenc}
\usepackage{amsmath}
\usepackage{amsbsy}
\usepackage{color}
\usepackage{rotating}
\usepackage[english]{babel}
\usepackage{multirow}

\def\be {\begin{equation}}
\def\ee {\end{equation}}
\def\bea {\begin{eqnarray}}
\def\eea {\end{eqnarray}}
\def\ba {\begin{align}}
\def\ea {\end{align}}

\newcommand{\dd}{{\rm d}}

\newcommand{\kappat}{\tilde{\kappa}}
\newcommand{\Od}{{\mathcal O}}

\newcommand{\lsim}   {\mathrel{\mathop{\kern 0pt \rlap
  {\raise.2ex\hbox{$<$}}}
  \lower.9ex\hbox{\kern-.190em $\sim$}}}
\newcommand{\gsim}   {\mathrel{\mathop{\kern 0pt \rlap
  {\raise.2ex\hbox{$>$}}}
  \lower.9ex\hbox{\kern-.190em $\sim$}}}

\begin{document}

\title{Cosmological future singularities in interacting dark energy models}

\author{Jose Beltr\'an Jim\'enez} \email{jose.beltran@cpt.univ-mrs.fr}
\affiliation{Aix Marseille Univ, Universit\'e de Toulon, CNRS, CPT, Marseille, France}
\author{Diego Rubiera-Garcia} \email{drgarcia@fc.ul.pt}
\affiliation{Instituto de Astrof\'{\i}sica e Ci\^{e}ncias do Espa\c{c}o, Faculdade de
Ci\^encias da Universidade de Lisboa, Edif\'{\i}cio C8, Campo Grande,
P-1749-016 Lisbon, Portugal}
\author{Diego S\'aez-G\'omez} \email{saez@ieec.uab.es}
\affiliation{Instituto de Astrof\'{\i}sica e Ci\^{e}ncias do Espa\c{c}o, Faculdade de
Ci\^encias da Universidade de Lisboa, Edif\'{\i}cio C8, Campo Grande,
P-1749-016 Lisbon, Portugal}
\affiliation{Institut de Ci\`{e}ncies de l'Espai, ICE/CSIC-IEEC, Campus UAB, Carrer de Can Magrans s/n, 08193 Bellaterra (Barcelona), Spain}
\author{Vincenzo Salzano} \email{enzo.salzano@wmf.univ.szczecin.pl}
\affiliation{Institute of Physics, University of Szczecin, Wielkopolska 15, 70-451 Szczecin, Poland}

\date{\today}

\begin{abstract}
The existence of interactions between dark matter and dark energy has been widely studied, since they can fit well the observational data and may provide new physics through such an interaction. In this work we analyze these models and investigate their potential relation with future cosmological singularities. We find that every future singularity found in the literature can be mapped into a singularity of the interaction term, that we call $Q$-singularity, where the energy flow between the dark components diverges. Furthermore, this framework allows to identify a new type of future singularity induced by the divergence of the first derivative of the dark energy equation of state parameter.
\end{abstract}


\maketitle

\section{Introduction}

Over the last decade, one of the main challenges in theoretical physics, and particularly in cosmology, refers to identifying the mysterious nature of the two dominant components that, according to observations and most of the theoretical frameworks, compose the universe, namely: the so-called dark energy and dark matter. While the latter is likely behaving nowadays as a pressureless fluid, the former should have an effective negative pressure in order to explain the accelerated expansion of the universe. Most of the dark energy models assume an extra and unknown field which would be the responsible for the accelerated expansion, but other realistic proposals include infrared modifications of General Relativity (for some reviews, see \cite{Nojiri:2010wj}). In any case, the main problem arises because of the large number of models, either extra fields or modified gravity, capable of explaining the observational data and leading to similar statistical evidence.

However, searching for models that include other effects may provide an alternative way of exploring the nature of the dark sector to, hopefully, allow for discriminating among the different theoretical models. In this sense, some proposals introduce the possibility of an interaction between dark matter and dark energy, which may reveal new features of both components (for a recent review see \cite{Wang:2016lxa}). Indeed, several interactions have been suggested, where basically the full Lagrangian contains a particular interaction term, including non-minimally coupled theories \cite{Koivisto:2005nr}. A different and more phenomenological way of exploring such a possibility goes directly through the field equations, where an interacting term, usually dubbed $Q$, is included in the continuity equations such that the total energy is conserved but a flow of energy exists between the two components, assuming that both behave like perfect fluids \cite{Zimdahl:2001ar}-\cite{Pereira:2008at}. Note that, despite the fact that these models are constructed from a phenomenological viewpoint, they can also be obtained from a variational principle \cite{Skordis:2015yra}. Moreover, a major motivation to explore dark couplings, besides the search for new physics, lies on the possibility of solving the coincidence problem, since some suitable interaction terms lead to attractor solutions with an order one ratio of dark matter and dark energy \cite{Zimdahl:2001ar,Boehmer:2008av}. The existence of such scenarios has been explored in different frameworks, from holographic dark energy to periodic universes or future singularities \cite{Nojiri:2005sr,Nojiri:2005sx}. An important issue that these models face is the potential presence of instabilities at early cosmological times \cite{He:2008si}. In any case, it is interesting to notice that some cosmological observations point out to the viability of some of these interactions \cite{Pereira:2008at}.

The required negative pressure for the dark energy component  has led to exploring models with an effective equation of state (EoS) that violates the null energy condition (NEC), $\rho+p>0$, usually called {\it phantom} \cite{phantom}. Violations of the NEC can be easily achieved with extra fields that can arise from modified gravity theories or high energy physics. However, violating the NEC is not harmless and, in fact, phantom models may give rise to divergences in some cosmological parameters occurring at a finite time, thus potentially inducing physical singularities that jeopardize the regular structure of the spacetime. Moreover, in many situations, these divergences additionally signal the presence of pathologies in the perturbations. In fact, a phantom cosmological evolution supported by minimally coupled scalar fields necessarily leads to the appearance of laplacian and/or gradient instabilities in the perturbations.  On the other hand, the so-called {\it Big Rip}, which is one of the most extensively studied future singularities, is characterized by a divergent scale factor at a finite time, called the {\it Rip} time, and this makes every binding structure in the universe eventually break apart. Other future singularities analyzed in the literature lead to regular spacetimes in the sense of geodesic completeness, but still may give rise to arbitrarily large tidal forces so that the passage of physical structures through the singularity is not free of peril. It is worth mentioning that the standard lore relies on quantum effects to tame all these divergences, since quantum corrections should become important when the Planck scale is reached \cite{Nojiri:2004ip}.

In addition to the theoretical appealing of these scenarios, it is interesting to note that an effective EoS for dark energy $w<-1$ is allowed by observational data (and, from some sources, even favoured \cite{Lazkoz:2006gp}) and, thus, a universe with a future singularity might be a plausible scenario for our universe. In this sense, the classification provided in Ref.~\cite{Nojiri:2005sx} and updated in Ref.~\cite{Fernandez-Jambrina:2014sga}, shows how each divergence, usually appearing in the scale factor and its derivatives, affects the universe expansion and its structures. Moreover, there are some alternative non-singular scenarios (understood as the absence of divergences in finite time) that may lead to the break of some structures, as the {\it Little Rip} \cite{Frampton:2011sp}, {\it Pseudo-Rip} \cite{PseudoRip} and {\it Little Sibling} \cite{Mariam}.

In this paper, we present an analysis of future singularities within the framework of interacting dark energy-dark matter models. The appearance of different singular cosmological scenarios in interacting models with variable cosmological constant and exotic quintessence fields has been analyzed in \cite{Chimento}. Here we will describe the interactions in an effective way through the continuity equations and we find that every singularity found so far in the literature can be mapped into a singularity of the interaction term $Q$, and so we dub it {\it $Q$-singularity}. This means that the flow of energy between the dark matter and dark energy components diverges at a finite time, inducing a divergence in the scale factor and/or its derivatives and leading to one of the future singularities analyzed in the literature. Furthermore, a new future singularity is analyzed, where the $Q$-singularity provides a divergence in the derivative of the EoS for dark energy. We then study some specific interacting terms and EoS for dark energy, analyzing those cases where the interaction diverges.

The paper is organized as follows: in section \ref{sec:II} a review on cosmological singularities is given. Section \ref{sec:III} is devoted to the introduction of the $Q$-singularities, where several ansatz are considered and some models reconstructed. In section \ref{sec:IV} some interacting terms, previously analyzed in the literature, are considered, where for some particular EoS for dark energy a $Q$-singularity occurs. Finally, section \ref{sec:V} contains the conclusions of the paper.

\section{Future singularities in Cosmology} \label{sec:II}

Assuming a homogeneous and isotropic universe at large scales, in compliance with the cosmological principle, the line element is given by Friedman-Lema\^itre-Robertson-Walker (FLRW) metric
\be \label{eq:FLRW}
ds^2=-dt^2+a(t)^2\left(dx^2+dy^2+dz^2\right),
\ee
where $a(t)$ is the scale factor and we have assumed spatially flat sections. For the matter sector we shall consider a perfect fluid, whose energy-momentum tensor is given by
\be \label{eq:pf}
T_{\mu\nu}=(\rho+p) u_{\mu} u_{\nu} + p g_{\mu\nu}\ ,
\ee
where $u_{\mu}$ is a normalized timelike  vector, $u_{\mu}u^{\mu}=-1$, and $\rho$ and $p$ are the energy density and pressure of the fluid, respectively. Under these assumptions, the Einstein equations, $G_{\mu\nu}(g)=\kappa^2 T_{\mu\nu}$, where $\kappa^2=8\pi G$ with $G$ the Newton's constant, yield the equations
\be \label{eq:Fried}
H^2=\frac{\kappa^2 }{3} \rho \quad{\rm and}\quad \dot{H}=-\frac{\kappa^2}{2}(\rho+p)\ .
\ee
These equations determine the background evolution of the Hubble parameter, $H \equiv \frac{\dot{a}}{a}$ (a dot denotes a time derivative), once the matter content (\ref{eq:pf}) is fully specified. The above equations can be combined to obtain the continuity equation $\dot{\rho}+3H(\rho+p)=0$, which can be trivially solved for a constant EoS, $w \equiv \frac{p}{\rho}$, as $\rho(t) \propto a^{-3(1+w)}$. Inserting this result into the Friedman equations (\ref{eq:Fried}) yields the well known result
\begin{equation} \label{eq:Hasol}
H=\frac{2}{3(1+w)(t-t_s)} \Rightarrow a(t) \propto (t-t_s)^{\frac{2}{3(1+w)}} \ ,
\end{equation}
where $t_s$ is some constant. Then one just needs to provide a specific EoS $w$ and Eq.(\ref{eq:Hasol}) gives a full solution. This immediately brings forward a potential problem: if $w<-1$ the above solution leads to an expanding universe such that at $t=t_s$ both the scale factor $a(t)$ and the energy density $\rho$ diverge. This is indeed the tip of a broader problem, namely: depending on the properties of the matter under consideration, some cosmological quantities may diverge at a finite time $t=t_s$.

In a curved space-time the trouble with singularities is very subtle and a broad literature deals with this problem from different perspectives. In this sense, the most rigorous and well accepted criterion about the nature of spacetime singularities relies on the concept of \emph{geodesic completeness}, namely, whether any null and time-like geodesic can be extended to arbitrarily large values of its affine parameter or not \cite{Geodesics,Wald}. In this sense, a number of theorems have been established to determine the conditions upon which a given spacetime contains a singularity \cite{Theorems}. These theorems give a precise mathematical formulation to the physically intuitive idea that, being null geodesics attached to the transmission of information and time-like geodesics to the free falling of idealized physical observers, in a well behaved spacetime nothing should suddenly cease to exist or emerge from nowhere.

Let us review the analysis of the geodesic equation when assuming a flat FLRW spacetime (\ref{eq:FLRW}). A geodesic curve $\gamma^{\mu}=x^{\mu}(\lambda)$, where $\lambda$ is the affine parameter, satisfies, in a coordinate system, the following equation \cite{Chandra,Wald}
\be \label{eq:geoeq}
\frac{\dd^2 x^\mu}{\dd \lambda^2}+\Gamma^\mu_{\alpha\beta}\frac{\dd x^\alpha}{\dd \lambda}\frac{\dd x^\beta}{\dd \lambda}=0 \ ,
\ee
where $\Gamma^\mu_{\alpha\beta}$ are the Christoffel symbols of the spacetime metric $g_{\mu\nu}$. This way, the geodesic equations (\ref{eq:geoeq}) become
\bea
\frac{\dd^2 t}{\dd \lambda^2}+Ha^2\delta_{ij}\frac{\dd x^i}{\dd \lambda}\frac{\dd x^j}{\dd \lambda}=0\, ,\label{tgeodesic}\\
\frac{\dd^2 x^i}{\dd \lambda^2}+2H\frac{\dd x^i}{\dd \lambda}\frac{\dd t}{\dd \lambda}=0 \label{xgeodesic}.
\eea
By using $H=\frac{\dot{a}}{a}=\frac{1}{a}\frac{\dd a/\dd \lambda}{\dd t/\dd\lambda}$, Eq. (\ref{xgeodesic}) can be rewritten as follows
\be
\frac{\dd}{\dd \lambda}\left(a^2\frac{\dd x^i}{\dd \lambda}\right)=0 \ ,
\ee
which gives
\be
\frac{\dd x^i}{\dd \lambda}=\frac{u^i_0}{a^2} \ ,
\ee
where $u^i_0$ are integration constants. Then, using the above result, Eq. (\ref{tgeodesic}) yields
\be
\left(\frac{\dd t}{\dd\lambda}\right)^2=\frac{\vert\vec{u}_0\vert^2}{a^2}+C_0 \ ,
\ee
where $C_0$ is another integration constant. We thus see that the geodesics will be regular (with a well defined tangent vector) as long as the scale factor remains regular. Hence, if the scale factor does not diverge and is non-vanishing (so the metric is regular), the 4-velocities of the geodesics remain regular and the spacetime will be said to be non-singular. If the scale factor diverges at some point, then the geodesics stop there and cannot go through it. As we have discussed above, it is important to notice that the geodesics are insensitive to divergences in the expansion rate $H$ or its derivatives if they do not correspond to a singular behavior of the scale factor. This will be the case of the types II, III and IV singularities below, where the scale factor remains finite while all the divergences only appear in its derivatives.

When any geodesic path cannot be indefinitely extended, one would be interested in understanding the underlying reason for that. Taking into account that in many spacetimes their geodesically incomplete character comes alongside the divergence of (some) curvature scalars, one might blame the presence of infinitely large tidal forces for the existence of incomplete paths. Therefore, a framework has been developed to determine the impact of tidal forces upon physical (extended) observers \cite{TK}, establishing the criteria of \emph{strong} singularities if the body is unavoidable destroyed as it crosses the divergent region, and \emph{weak} in case it could retain its identity, i.e., its finite extended nature. There are two broadly used criteria (known as Tipler and Krolak criteria) to classify singularities as weak or strong  according to the convergence of the following integrals:
\bea
T(u)&\equiv&\int_0^\lambda \dd\lambda' \int_0^{\lambda'} \dd \lambda''R_{ij}u^i u^j,\\
K(u)&\equiv&\int_0^\lambda \dd{\lambda'} ''R_{ij}u^i u^j.
\eea
where $u^i$ is the 4-velocity of the geodesic towards the singularity and $R_{ij}$ the components of the Ricci tensor. From these expressions it is clear that a spacetime containing a divergence in (some of) the curvature scalars can still be regular according to the above criteria.

Equipped with the two tools described above (geodesic completeness and weak/strong singularities), a number of future singularities have been found and studied in detail in the literature:
\begin{itemize}
  \item Type I (``Big Rip singularity"): For $t \rightarrow t_s$, $a \rightarrow \infty$, $\rho \rightarrow \infty$ and $ \vert p \vert  \rightarrow \infty$. This case yields incomplete null and time-like geodesics \cite{TI}. Thus it represents a genuine space-time singularity.
  \item Type II (``Sudden singularity"): For $t \rightarrow t_s$, $a \rightarrow a_s$, $\rho \rightarrow \rho_s$ and $\vert p \vert \rightarrow \infty$. Geodesics are complete and observers are not necessarily crushed (weak singularity \cite{TII,barrowIV}).
  \item Type III (``Big Freeze singularity"): For $t \rightarrow t_s$, $a \rightarrow a_s$, $\rho \rightarrow \infty$ and $ \vert p \vert  \rightarrow \infty$. Geodesically complete solutions, which can be either strong or weak \cite{TIII}
  \item Type IV (``Generalized Sudden singularity"): For $t \rightarrow t_s$, $a \rightarrow a_s$, $\rho \rightarrow \rho_s$ and $ \vert p \vert  \rightarrow p_s$ but second and higher derivatives of the Hubble parameter $H$ diverge. Geodesics are complete and the singularity is weak \cite{barrowIV,TIV}.
  \item Type V (``$w$-singularity"): For $t \rightarrow t_s$, $a \rightarrow \infty$, $\rho \rightarrow 0$ and $ \vert p \vert  \rightarrow 0$ but the equation of state $w \rightarrow \infty$. These singularities are weak as well \cite{TV}.
\end{itemize}
This list summarizes the current knowledge on cosmological singularities in the literature, where the divergent quantities are identified. In this classification it is common to implicitly assume that the usual Friedman equations hold, so that the singularities originate from some exotic properties of the matter sector (e.g., violation of the NEC). However, it is worth pointing out that some scenarios might lead to the appearance of cosmological future singularities for non-exotic matter fields, e.g dust or radiation fluids. This is the case, for instance, in some Born-Infeld inspired theories \cite{singularBI}, $f(R)$ gravities \cite{Nojiri:2003ft} or modified gravity theories formulated in generalized Weyl geometries \cite{Jimenez:2016opp}.

\section{Q-singularities} \label{sec:III}

In this section we shall further specify the setup discussed in the previous section and consider that the matter sector comprises non-relativistic dark matter, with $w_{m}=0$, and dark energy, with $w_{DE}  \equiv p_{DE}/\rho_{DE} \neq 0$ a certain function of time. These two components will be assumed to interact according to \cite{Kodama}
\begin{equation}
\nabla_{\mu} {T^{\mu(m)}}_{\nu}=Q_{\nu} \hspace{0.1cm};\hspace{0.1cm} \nabla_{\mu} {T^{\mu(DE)}}_{\nu}=-Q_{\nu} \ ,
\label{Inteq}
\end{equation}
where the $4$-vector $Q_{\nu}$ governs the stress-energy transfer between the two dark components. The choice of $Q_\nu$ will determine the specific model under consideration. A natural choice is to assume that this vector lies within the space spanned by the 4-velocities of dark matter and dark energy. While this is important at the level of the perturbations, the fact that dark energy and dark matter are usually assumed to share a common rest frame on large scales makes it irrelevant for the homogeneous evolution\footnote{In models of moving dark energy \cite{MDE}, the two dark components can have a relative motion even at the background level so that more general interactions could be envisioned. We will not consider these scenarios here and, in any case, if dark energy interacts with dark matter, they are expected to have a common rest frame at large scales.}. In other words, dark matter and dark energy have the same background 4-velocity $u_\nu$ and, thus, the interaction term simply becomes $Q_{\nu}=Q u_{\nu}$. With these considerations in mind, the Eqs.(\ref{Inteq}) read
\begin{eqnarray}
\dot{\rho}_m+3H\rho_m&=&Q(t)\ , \label{Eqbis1} \\
\dot{\rho}_{DE}+3H(1+w_{DE})\rho_{DE}&=&-Q(t) \label{Eqbis3} \ .
\end{eqnarray}
and which is nothing but the field equations of the matter sector. In these equations, $Q(t)$ accounts for the energy exchange rate between the two dark sectors, so that $Q(t)>0$ implies a transfer of energy from the dark matter sector to the dark energy one and the other way around for $Q(t)<0$. Although we have added an interaction between the two dark components, the modification is such that the gravitational field equations remain unchanged
\begin{eqnarray}
H^2&=&\frac{\kappa^2}{3}\left(\rho_m+\rho_{DE}\right)\label{FLRWeq2}  \ , \\
\dot{H}&=&-\frac{\kappa^2}{2} \left[\rho_m+\rho_{DE}+p_{DE}\right] \label{FLRWeq3} \ .
\end{eqnarray}
These two equations are consistent with the Bianchi identities as well as with the modified continuity equations. We can combine the two gravitational equations to express the dark matter and dark energy densities in terms of the Hubble expansion rate and the dark energy equation of state as
\begin{eqnarray}
\rho_{DE}&=&-\frac{1}{w_{DE}\kappa^2} \left(3H^2 + 2\dot{H} \right) \\
\rho_{m}&=&\frac{1}{w_{DE}\kappa^2} \left[3(1+w_{DE})H^2 + 2\dot{H} \right].
\end{eqnarray}
We can now combine these equations with the matter ones (\ref{Eqbis1}), (\ref{Eqbis3}) so that we finally obtain an expression for $Q(t)$ as
\begin{eqnarray}\label{Eqbis2}
Q(t) &=& \frac{1}{\kappa^2 w_{DE}} \left[9(1+w_{DE})H^3+6(2+w_{DE})H\dot{H} \right. \nonumber \\
&+&2\ddot{H}- \left.  \frac{\dot{w}_{DE}}{w_{DE}} \left( 3 H^2 + 2 \dot{H}\right) \right].
\end{eqnarray}
Hence, by assuming a particular cosmological evolution $H=H(t)$, the corresponding interacting term $Q(t)$ can be obtained.
In the above expression for $Q(t)$ we can see that a future singularity implying a divergence in $H$, $\dot{H}$ or $\ddot{H}$ will typically induce a divergence in $Q$. Moreover, also divergences in the equation of state will also lead to a divergent interaction. As a novel feature, we additionally find that a divergence in $\dot{w}_{DE}$ gives rise to a singular interaction as well. Since the background field equations do not involve the derivatives of the equation of state parameter, this type of divergences is expected to be harmless for the homogeneous evolution, but the perturbations might be sensitive to them since the adiabatic sound speed in a barotropic fluid is given by
\be
c_s^2=\frac{\dot{p}}{\dot{\rho}}=w\left(1+\frac{\rho\dot{w}}{\dot{\rho}w}\right),
\ee
and, therefore, a divergence in $\dot{w}_{DE}$ could induce a divergence in the sound speed of dark energy. Here we are mainly interested in how the different types of singularities can be mapped into a divergence of $Q$ so that we will not explore any further these potential physical implications, which, in addition, would require a full covariant formulation to study the perturbations.

We will end our general treatment by giving an alternative relation between the interaction term and the Hubble expansion rate that will complement Eq.(\ref{Eqbis2}). For that, we first note that the continuity equations (\ref{Eqbis1}) and (\ref{Eqbis3}) can be solved for the energy densities given arbitrary expansion rate, dark energy equation of state and interaction term. The solutions can then be expressed as
\bea
\rho_m&=&\bar{\rho}_m\left(1+\int\frac{Q}{\bar{\rho}_m}\dd t\right)\\
\rho_{DE}&=&\bar{\rho}_{DE}\left(1-\int\frac{Q}{\bar{\rho}_{DE}}\dd t\right),
\eea
where $\bar{\rho}_m$ and $\bar{\rho}_{DE}$ denote the homogeneous standard solutions in non-interacting models, i.e.,
\bea
\bar{\rho}_m&=&\rho_m^0a^{-3}\\
\bar{\rho}_{DE}&=&\rho_{DE}^0\int e^{-3\int H(1+w_{DE})\dd t}\dd t,
\eea
with $\rho_m^0$ and $\rho_{DE}^0$ integration constants. We can then insert these solutions into the Friedman equation (\ref{FLRWeq2}) to obtain:
\be
H^2-\frac{\kappa^2}{3}\big(\bar{\rho}_m+\bar{\rho}_{DE}\big)=
\frac{\kappa^2}{3}\left(\bar{\rho}_m\int\frac{Q}{\bar{\rho}_m}\dd t-\bar{\rho}_{DE}\int\frac{Q}{\bar{\rho}_{DE}}\dd t\right).
\label{eqFbis1}
\ee
Analogously, the second gravitational equation (\ref{FLRWeq3}) can be expressed as
\bea
&\dot{H}&+\frac{\kappa^2}{2}\big[\bar{\rho}_m+(1+w_{DE})\bar{\rho}_{DE}\big]\label{eqFbis2}\\
&=&-\frac{\kappa^2}{2}\left[\bar{\rho}_m\int\frac{Q}{\bar{\rho}_m}\dd t-(1+w_{DE})\bar{\rho}_{DE}\int\frac{Q}{\bar{\rho}_{DE}}\dd t\right].\nonumber
\eea
In these expressions we have explicitly separated the usual gravitational equations in the absence of interactions (LHS) from the modifications coming from the interactions (RHS).  We can easily see that a divergence in $Q$ can give rise to divergences in $H$ or $\dot{H}$, but  one can have a singular  interaction $Q$ while $H$ and $\dot{H}$ remain finite. This is so because $Q$ only enters the above expressions inside the integrals, which can improve the smoothness of $Q$. In order to obtain Eq. (\ref{Eqbis2}) we had to take derivatives of the equations and this can introduce additional divergences that might lack physical relevance (like the singularities originating from $\dot{w}$ that are expected to have effects only at the perturbations level). However, expressions (\ref{eqFbis1}) and (\ref{eqFbis2}) are obtained after solving the continuity equations and, thus, divergences appearing there have a more direct physical relevance.

In this work we are interested in knowing under which circumstances the interacting term yields future singularities on the cosmological background evolution. To this end we shall split our analysis into constant and time-dependent EoS for the dark energy component in the following.

\subsection{$w_{DE}=$constant}

This choice removes the $\dot{w}_{DE}$ contribution in Eq.(\ref{Eqbis2}). Let us parameterize the Hubble factor as
\be \label{eq:paramH}
H(t)=A+\frac{2}{3t}+B(t_s-t)^{\alpha} \ ,
\ee
where $A$ and $B$ are some constants and $\alpha \neq 0$ controls the type of singularity at the time $t_s$. Note that such parametrization is assumed in such a way that the matter dominated epoch is recovered asymptotically while at late-times the cosmological constant term should dominate together with the last term in (\ref{eq:paramH}), which denotes deviations from $\Lambda$CDM model. For the appropriate value of $\alpha$, the last term will eventually dominate when approaching the singularity, if it occurs. From the parametrization (\ref{eq:paramH}) one obtains the dominant term for the scale factor $a(t)$ as
\be
a(t)=a_0t^{2/3} e^{\left[At+B(t_s-t)^{\alpha} \left(\frac{t-t_s}{1+\alpha} \right)\right] } \ ,
\ee
if $\alpha \neq -1$ and
\be
a(t)=a_0t^{2/3} \frac{e^{At}}{t_s-t} \ ,
\ee
if $\alpha =-1$. In these expressions $a_0$ is an integration constant. We can now expand $Q(t)$ as given in (\ref{Eqbis2}) around $t=t_s$ with the parametrization (\ref{eq:paramH}) to obtain
\be
Q(t) \simeq (t-t_s)^{\alpha-2} \ ,
\ee
which diverges for $\alpha<2$. We will proceed by splitting the analysis in different subcases for $\alpha$ to identify the type of future singularity and its relation to the divergence of $Q(t)$.

\begin{itemize}
    \item $\alpha=-1$: At $t=t_s$ one has
    \begin{eqnarray}
    a(t) &\simeq& a_0 \frac{t^{2/3} e^{At}}{t_s-t} \rightarrow \infty \\
    H(t) &\simeq& \frac{B}{t_s-t} \rightarrow \infty \ .
    \end{eqnarray}
    According to the classification introduced in section \ref{sec:II}, the divergence of both the scale factor and the Hubble parameter is distinctive of a \emph{Big Rip} singularity, which prevents the completeness of geodesics. Note that it is the divergence of $H$ that induces the one of $Q(t)$ at $t=t_s$.
  \item $-1<\alpha<0$: At $t=t_s$ one has
  \begin{eqnarray}
  a(t) &\simeq& a_0 t^{2/3} e^{At_s}=a_s \\
  H(t) &\simeq& \frac{B}{(t_s-t)^{\vert x \vert}} \rightarrow \infty \ .
  \end{eqnarray}
  According to the classification introduced in Sec.\ref{sec:II}, this is a type III \emph{Big Freeze} singularity.
  \item $0<\alpha<1$: At $t=t_s$ we have
  \begin{eqnarray}
  a(t) &\simeq& t^{2/3} e^{At}=a_s \\
  H(t) &\simeq& A \ ,
  \end{eqnarray}
  and thus both are finite, corresponding to an asymptotically de Sitter space. However, the first derivative of the Hubble parameter, $\dot{H} \simeq \frac{B}{(t_s-t)^{\vert \alpha-1 \vert}}$, diverges at $t=t_s$, which is the term in (\ref{Eqbis2}) related to the divergence of $Q(t=t_s) \rightarrow \infty$. This is a \emph{Sudden} singularity.
  \item $1<\alpha<2$: At $t=t_s$ one has
  \begin{eqnarray}
  a(t) &\simeq& t^{2/3}e^{At} =a_s \\
  H(t) &\simeq& A \\
  \dot{H} &\simeq& -\frac{2}{3t^2} \ ,
  \end{eqnarray}
  and thus all of them finite. However, second time derivatives of $H$ appearing in Eq.(\ref{Eqbis2}) do diverge, $\ddot{H} \simeq \frac{B\alpha (\alpha-1)}{(t_s-t)^{\vert \alpha \vert}} \rightarrow \infty$. This is a \emph{Generalized Sudden} singularity.
\end{itemize}
It is also worth pointing out that in those cases with $\alpha <-1$, at $t=t_s$ one has

\begin{eqnarray}
a(t) &\simeq& a_0 \frac{t^{2/3}}{(t_s-t)^{\vert 1+\alpha \vert}}  e^{-\frac{B}{\vert 1 + \alpha \vert}} \rightarrow 0 \\
H(t) &\simeq& \frac{B}{(t_s-t)^{\vert \alpha \vert}} \rightarrow \infty \ ,
\end{eqnarray}
which is a Big Bang type singularity.

The analysis above shows that the four classes of future singularities introduced by Nojiri, Odintsov and Tsujikawa in Ref.\cite{Nojiri:2005sx} can be understood to be just particular cases of singularities of the function $Q(t)$ at $t=t_s$, where each type of divergence in the scale/Hubble factor or its derivatives comes from a different degree of divergence of $Q(t)$. In this sense, if $\alpha \geq 2$ the function $Q(t)$ is finite and no future singularity emerges.

\subsection{$w_{DE} \equiv w_{DE}(t)$}

Next we will extend our analysis to a non-constant dark energy EoS. Let us parameterize it around $t=t_s$ as
\be \label{eq:omegaeos}
w_{DE}\simeq w_{s}+(t_s-t)^{\beta} \ ,
\ee
with $w_{s}$ some constant. A glance at Eq.(\ref{Eqbis2}), in combination with Eq.(\ref{eq:omegaeos}), tells us that if the parameter $\beta<1$ the term $\dot{w}_{DE}/w_{DE}^2$ will diverge. There are indeed several cases to be analyzed separately:

\begin{itemize}
  \item If $-1<\beta<0$ then we have
    \begin{eqnarray}
    w_{DE}(t) &\simeq& \frac{1}{(t_s-t)^{\vert \beta \vert}} \rightarrow \infty \\
    Q(t) &\simeq& \frac{\beta}{(t_s-t)^{\vert 1-\beta \vert} \left(w_s+(t_s-t)^{\beta} \right)^2} \rightarrow \infty \ .
    \end{eqnarray}
    Thus in this case we have a divergence in the function $w_{DE}$, which induces the one of $Q(t)$. These are $w_{DE}$-singularities, which correspond to the  type V singularities discussed in Sec.\ref{sec:II}.
  \item If $0<\beta<1$ then at $t=t_s$ one has
  \begin{eqnarray}
  w_{DE}(t) &\rightarrow& w_s \\
  \dot{w}_{DE} &\simeq& \frac{1}{(t_s-t)^{\vert \beta -1 \vert}} \rightarrow \infty \\
  Q(t) &\rightarrow& \infty \ .
  \end{eqnarray}
This is a $\dot{w}_{DE}$-singularity, i.e., it is this contribution in (\ref{Eqbis2}) the responsible of inducing a divergence in $Q(t)$ (while $w$ remains finite). As this scenario is not included into the five types of future singularities discussed in Sec.\ref{sec:II}, we call them \emph{type VI singularities}. Assuming finiteness of both $a(t)$ and $H(t)$ in this case (otherwise one would end up into one of the type I-IV singularities above), it is easy to see that the corresponding spacetimes are geodesically complete due to the finiteness of the geodesic equations (\ref{tgeodesic}) and (\ref{xgeodesic})  (recall the discussion of Sec.\ref{sec:II}) and, likewise their $w_{DE}$-singularities partners, these type-VI singularities are weak. As discussed above, although this type of singularity does not represent any divergence for the homogeneous cosmological evolution, it might induce divergences in the perturbations through the adiabatic sound speed of dark energy. Finally, it would remain to check if this type of singularity can happen for a realistic physical model.
  \item If $\beta<-1$ then the term $\dot{w}_{DE}/w_{DE}^2$ yields a finite contribution to $Q(t)$ in Eq.(\ref{Eqbis2}). In such a case one would need to consider the behaviour of the other terms in Eq.(\ref{Eqbis2}), thus obtaining again the types I-IV of singularities.
\end{itemize}

From the analysis above it is clear that the known five types of future singularities (and the new type VI found here) can be seen just as particular cases of the divergence of the energy flow between the dark components, that is, {\it $Q$-singularities}. In the next section we shall review some particular interacting models and discuss their relation with $Q$-singularities.

\section{Dark energy/dark matter couplings revisited}  \label{sec:IV}
In this section we will consider some specific models for interacting dark energy and dark matter where we can explicitly show the appearance of different types of future cosmological singularities induced by the interacting terms. 

\subsection{$Q=\zeta H \rho_{DE}$} \label{sec:IVA}

Over the last years, couplings between dark energy and dark matter have been widely explored in the literature where the effective $Q$-term is taken to be proportional to the energy density, either the dark matter or the dark energy or a combination of both, i.e. (see e.g. \cite{Boehmer:2008av,Chimento:2007yt,Gavela,Salvatelli})
\be
Q=\zeta H \rho\ ,
\label{4.1}
\ee
where $\zeta$ is a constant to be determined by the observations and $\rho$ can be an arbitrary combination of the dark matter and dark energy energy densities. Then, let us explore the possibility of the existence of a $Q$-singularity when considering this type of interactions.

\begin{figure*}[htbp]
\centering
\includegraphics[width=9.5cm]{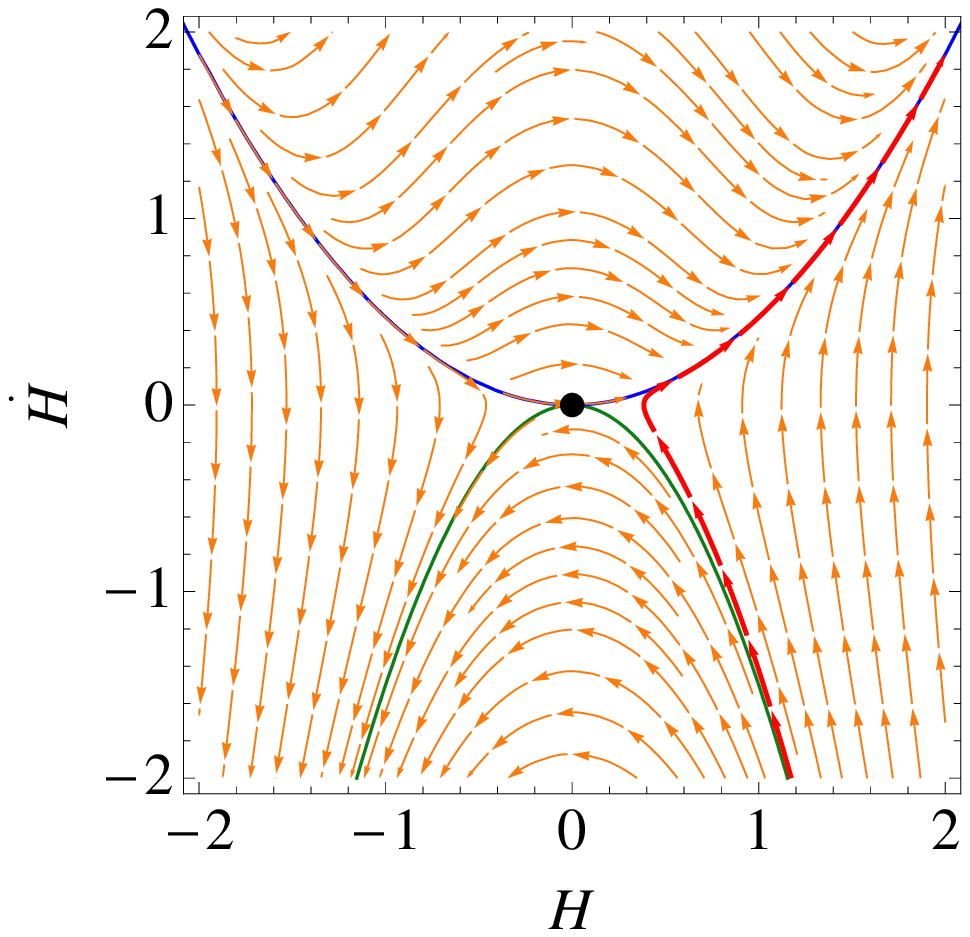}\hspace{0.5cm}
\includegraphics[width=6.5cm]{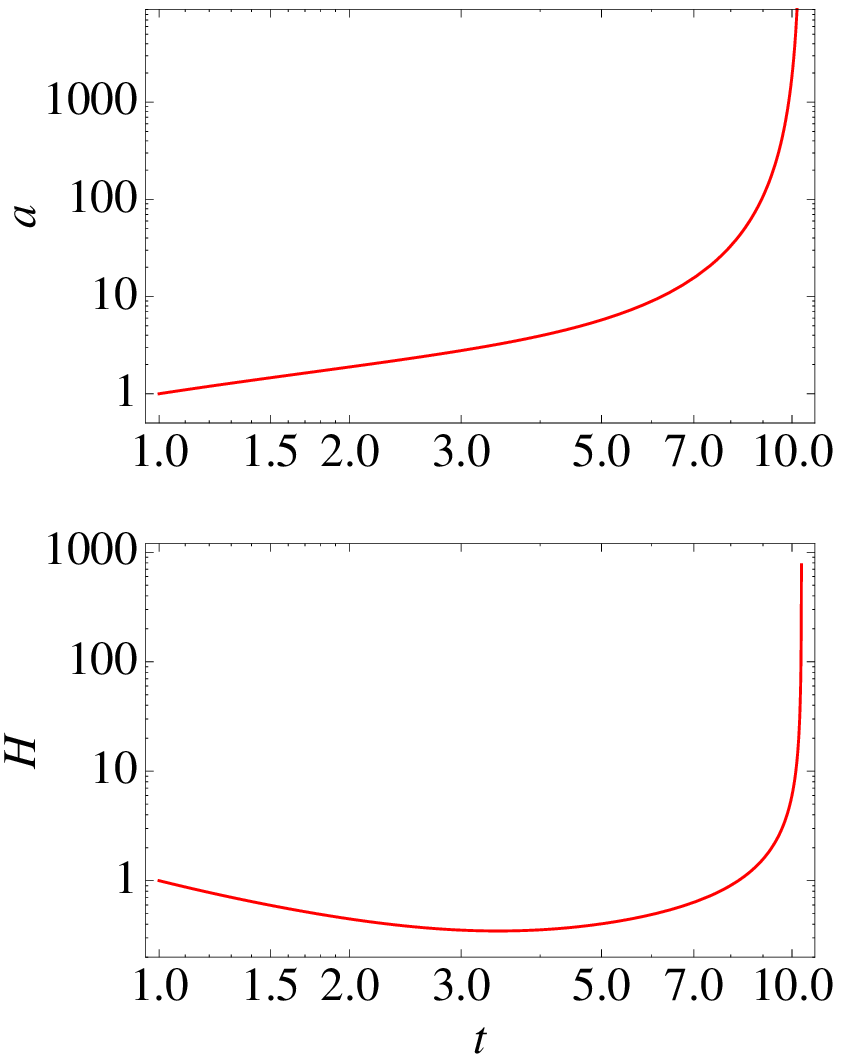}
\caption{In this figure we plot the phase map (left panel) and numerical solutions (right panels) of the equation (\ref{4.3}) corresponding to the interaction $Q=\zeta H\rho_{DE}$  and with the parameters set to $\zeta=-1$ and $w_{DE}=-0.98$. We also show the solution for a matter dominated universe (green line) and the asymptotic solution with $\dot{H}=\lambda^{-1} H^2$ (blue) discussed in the main text. The black dot denotes the unstable Minkowski critical point. The red trajectory corresponds to the solution with the scale factor and Hubble expansion rate depicted in the right panel (where we have normalized to their initial values). This solution transits from matter domination to dark energy domination eventually approaching a Big Rip singular solution even though dark energy is not phantom, thus confirming the analytical findings discussed in the main text.}\label{plotBigRip}
\end{figure*}


Firstly, we will focus on the case when the interacting term is proportional to the dark energy density $Q=\zeta H \rho_{DE}$. In such a case, by combining the FLRW equations (\ref{eq:Fried}), the interacting term $Q$ can be expressed as follows:
\be
Q=\zeta H\frac{3H^2+2\dot{H}}{\kappat^2}\
\label{4.2}
\ee
where $\kappat^2=-w_{DE}\kappa^2$. From this expression is clear that a possible $Q$-singularity can only induce a {\it Big Rip} or a {\it Sudden} singularity, depending on the magnitude that diverges in (\ref{4.2}), $H$ and/or $\dot{H}$. It is important to note that these are only potential singularities that might occur. In particular, Big Rip singularities involve a divergence of the scale factor as well, and this cannot be directly inferred from the expression for $Q$, but we need to look at the corresponding solutions. We can reduce the cosmological evolution within this scenario to a second order differential equation for the Hubble expansion rate by combining expressions (\ref{Eqbis2}) and (\ref{4.2}) to obtain:
\bea
&&2\ddot{H}+9(1+w_{DE})H^3+6(2+w_{DE})H\dot{H} \nonumber \\
&-&\frac{\dot{w}_{DE}}{w_{DE}} \left( 3 H^2 + 2 \dot{H}\right)+\zeta H\left(3H^2+2\dot{H}\right)=0\ .
\label{4.3}
\eea
From this equation we can analyse the behaviour of $H(t)$ and, thus, study the presence of future singularities. The remaining cosmological quantities, i.e., the energy densities, are  algebraically related to the solutions of the above equation and, therefore, the cosmological evolution is completely determined by (\ref{4.3}). For simplicity, from now on we will assume constant equation of state for dark energy $\dot{w}_{DE}=0$. It is easy to see that the only critical point in that case is the Minkowski solution with $H=0$, which is unstable (see Fig.\ref{plotBigRip}). Moreover,  although the equation is non-linear, it is easy to obtain exact solutions by taking advantage of its time rescaling invariance. This motivates to look for solutions with $\dot{H}=\lambda^{-1} H^2 $ with $\lambda$ some dimensionless parameter. These solutions lead to the usual cosmological evolution given by
\be
H=\frac{\lambda}{t_s-t}\quad\Rightarrow\quad a(t)\propto \vert t_s-t\vert^{-\lambda}\ ,
\ee
with $t_s$ some reference time. When inserting this ansatz into (\ref{4.3}) we obtain the following equation for $\lambda$:		
\be
\big(3+2\lambda^{-1}\big)\Big[3(1+w_{DE})+\zeta+2\lambda^{-1}\Big]=0
\ee
with two branches of solutions, namely $\lambda=-2/3$ and $\lambda=-2/(3(1+w_{DE})+\zeta)$. The first branch corresponds to a matter dominated universe (which is unstable in the presence of the dark energy component), while the second branch corresponds to a universe where either the dark energy or the interaction term dominates. We see that, in that branch, the effective equation of state is given by $w_{\rm eff}=w_{DE}+\zeta/3$. Thus, an interaction term with $\zeta<-3(1+w_{DE})$ (transfer from dark matter to dark energy) can induce an effective phantom behaviour leading to a type I Big Rip singularity even if dark energy satisfies the null energy condition.  This can be easily understood from the dark energy conservation equation which, with the interacting term under consideration can be written as
\be
\dot{\rho}_{DE}+3H\left(1+w_{DE}+\frac13 \zeta\right)\rho_{DE}=0
\ee
where we see that dark energy acquires the aforementioned effective equation of state determining the fate of the cosmological evolution. In Fig. \ref{plotBigRip} we show the phase map corresponding to the cosmological evolution of this model and a particular singular solution where our analytical results are also numerically confirmed.

\begin{figure}[!h]
\centering
\includegraphics[width=8cm]{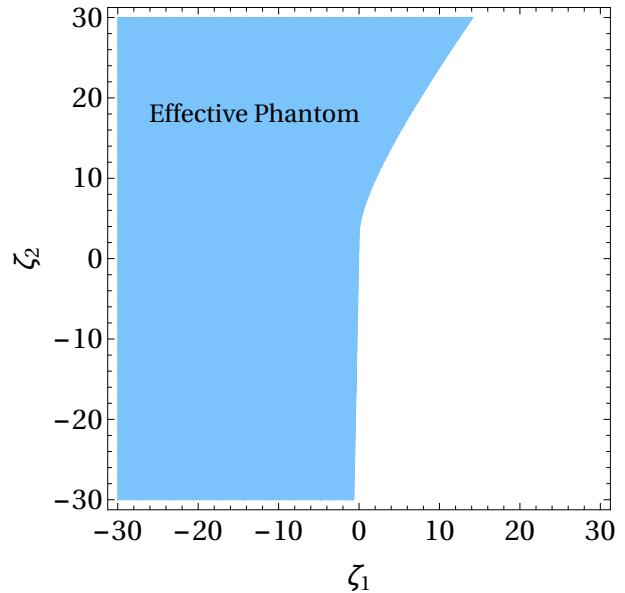}
\caption{In this plot we show the region in the parameter space where an interaction of the form $Q=H\Big(\zeta_1 \rho_{DE}+\zeta_2\rho_m\Big)$ can lead to a future Big Rip singularity characterized by an effective phantom behaviour even if dark energy satisfies the null energy condition (we have taken $w_{DE}=-0.98$).}\label{generalinteraction}
\end{figure}

Although the above results have been obtained for an interaction fully determined by the dark energy component, our findings are completely general for the type of interactions under consideration. In order to show that, let us now consider the more general interaction term given by a linear combination of $\rho_{DE}$ and $\rho_m$, i.e., $Q=H\Big(\zeta_1 \rho_{DE}+\zeta_2\rho_m\Big)$. By proceeding analogously, we find the following expression for $Q$
\be
Q=\frac{H}{\kappat^2}\Big[3\Big(\zeta_1-(1+w_{DE})\zeta_2\Big)H^2+2(\zeta_1-\zeta_2)\dot{H}\Big] \ .
\ee
For this general case we will also have solutions of the form $\dot{H}=\lambda^{-1}H^2$, where $\lambda$ is now given by
\begin{align}
\lambda^{-1}_{\pm}=&-\frac14\Big[3(2+w_{DE})+\zeta_1-\zeta_2\nonumber\\
&\pm\sqrt{(3w_{DE}+\zeta_1)^2+\zeta_2(6w_{DE}-2\zeta_1+\zeta_2)}\Big]\ .
\end{align}
We can see from this expression that we will also have future Big Rip singularities very much like in the previous case analyzed in detail. In this case the effective equation of state is given by $w_{\rm eff}=-(1+\frac{2}{3\lambda})$ and has a more complicated dependence on the model parameters, but it is easy to see that it can also lead to an effective phantom behaviour even if dark energy satisfies the null energy condition. In the limit of small couplings $\vert\zeta_{1,2}\vert\ll 1$ we obtain
\bea
w_{\rm eff}^+&\simeq& w_{DE}+\frac13\zeta_1\nonumber\\
w_{\rm eff}^-&\simeq& -\frac13\zeta_2 \ ,
\eea
for each corresponding branch. We see that the second branch can never lead to a Big Rip for small couplings $\vert\zeta_{1,2}\vert \ll 1$ because the effective equation of state is $w_{\rm eff}^-\sim\Od(\zeta_2)$. The first branch however, can give a Big Rip singularity for a non-phantom dark energy component if its equation of state is close to -1 and the interaction satisfies $\zeta_1<-3(1+w_{DE})$. Thus, if the interactions are small, only a negative $\zeta_1$ can induce an effective phantom behaviour. However, if the interactions are allowed to be larger, also a wide range of values of $\zeta_2$ and positive values of $\zeta_1$  can lead to future Big Rip singularities. In Fig. \ref{generalinteraction} we show the region in the parameter space where there is an effective phantom behaviour in the general case.

\begin{figure*}[!ht]
\includegraphics[width=9.2cm]{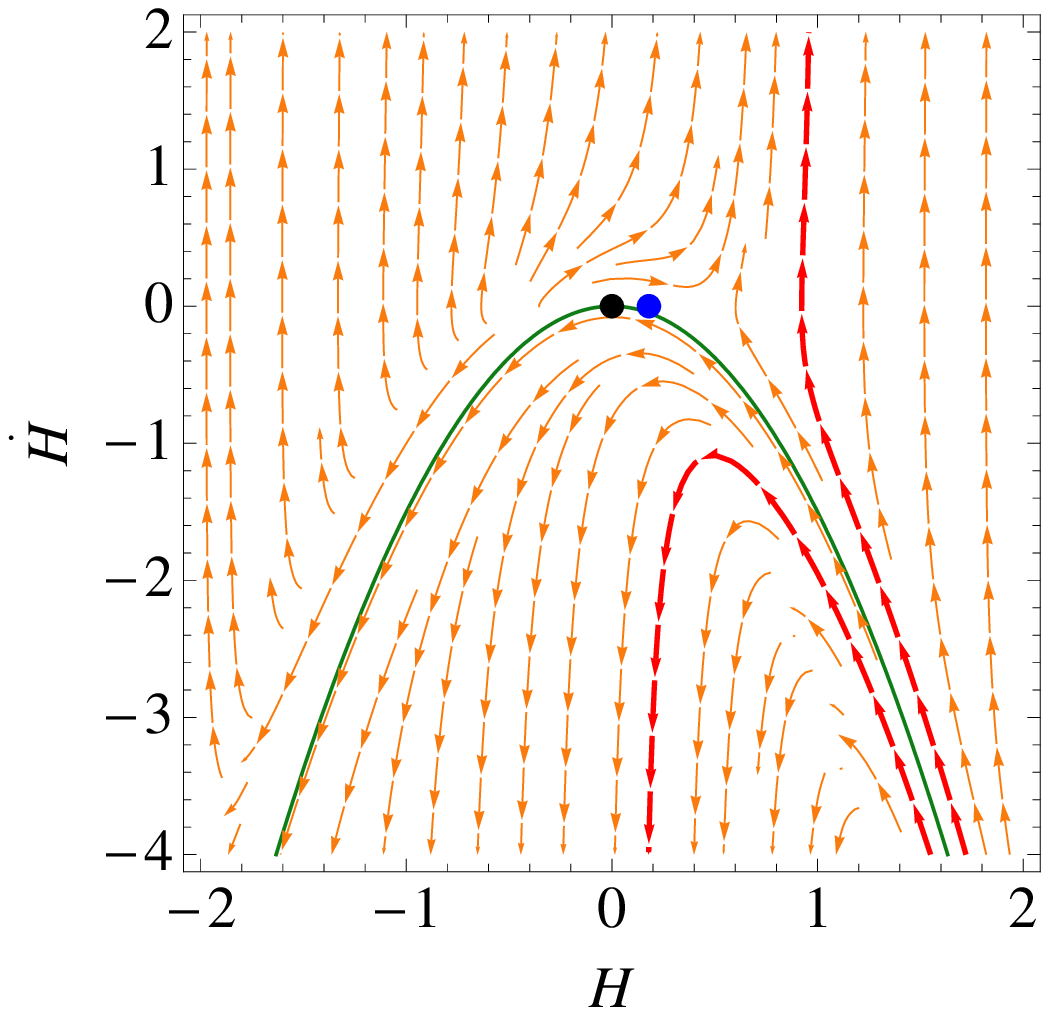}\hspace{0.5cm}
\includegraphics[width=6.5cm]{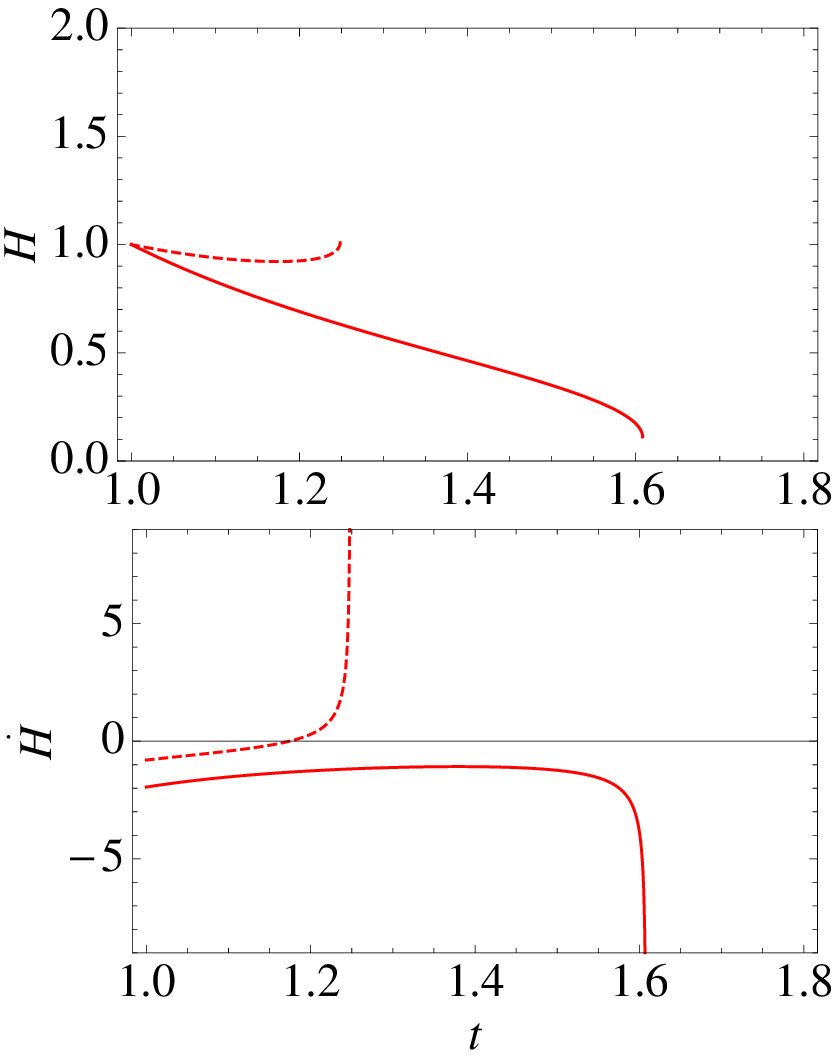}\\
\vspace{1cm}
\includegraphics[width=9.2cm]{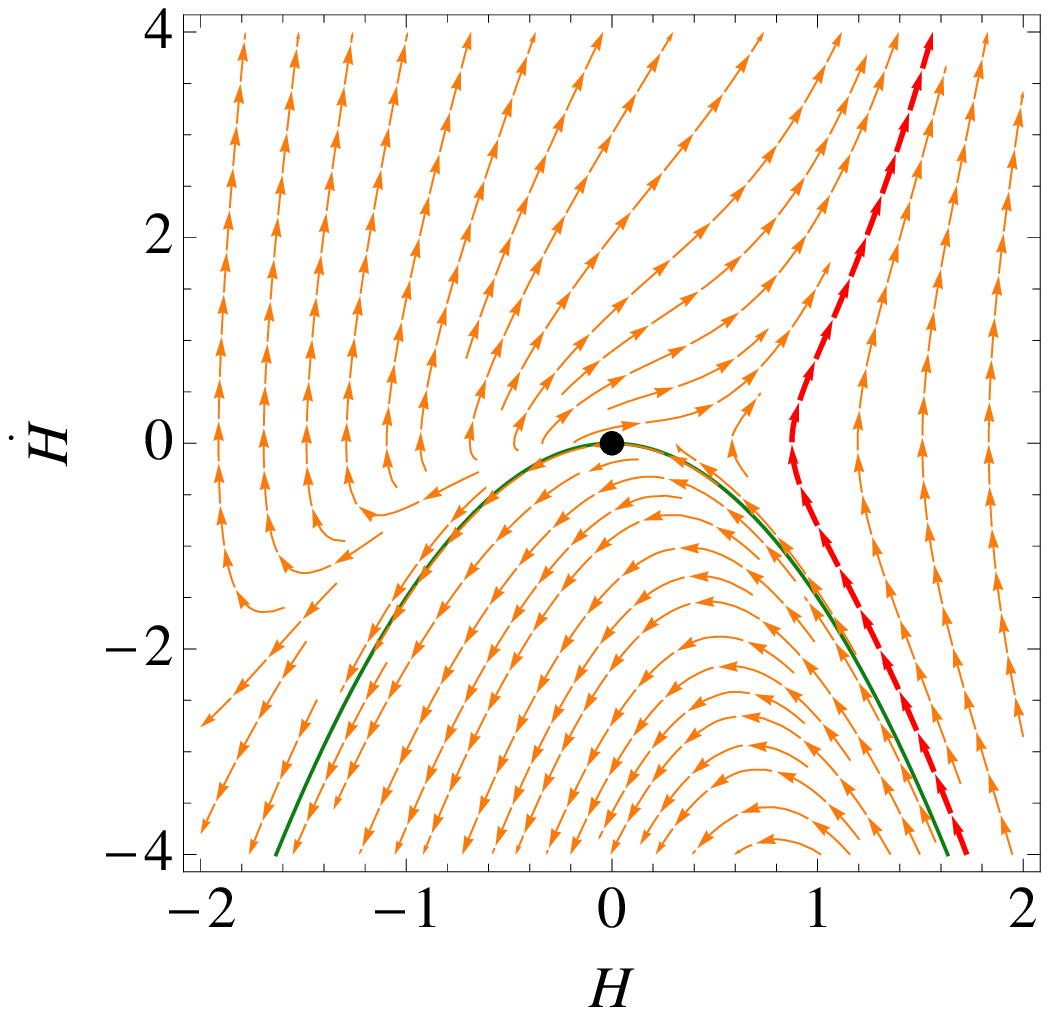}\hspace{0.5cm}
\includegraphics[width=6.5cm]{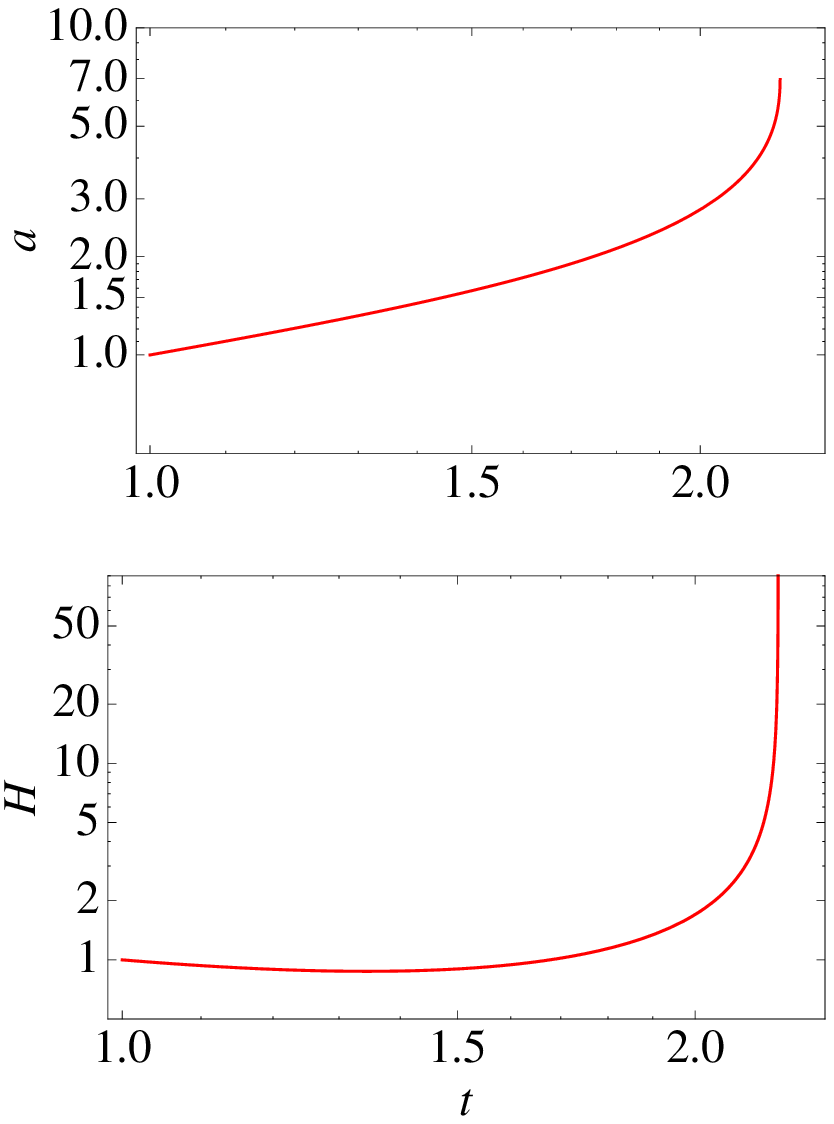}
\caption{In this plot we show the cosmological evolution for the interacting model $Q=\Gamma \rho_{DE}^n$. We have taken $w_{DE}=-0.98$ and $\Gamma=-1$ to show the presence of future singularities for non-phantom dark energy. In the upper panels we have taken $n=3$, which corresponds to the case exhibiting a future Sudden singularity with the asymptotic solution $H\simeq C(t_s-t)^p+H_s$ as discussed in the main text. In the upper left panel we show the phase map and indicate the matter dominated universe (green line), the critical points corresponding to Minkowski (black) and de Sitter (blue), both of which are unstable, and two trajectories whose numerical solutions are also shown in the upper right panels. We can see that most solutions asymptotically give a finite value of $H$ while $\dot{H}$ diverges, going either to $+\infty$ (dashed curves) or $-\infty$ (solid curves). These numerical solutions are in agreement and confirm the analytical results discussed in the main text. Notice also that there are solutions that remain close to the matter dominated universe but eventually evolve towards the singularity. This shows that there can indeed be a transition from matter domination to dark energy domination with a future singularity. Similarly, in the lower left panel we show the phase map for the case with $n=5/3$ that corresponds to a future Big Freeze singularity where the Hubble expansion rate (and its derivatives) diverges while the scale factor remains finite, as can bee explicitly seen in the numerical solutions given in the lower right panels, again confirming our analytical results.}\label{phasemapSudden}
\end{figure*}

\subsection{$Q=\Gamma\rho_{DE}^n$}

In the previous subsection we have shown that the most widely used interaction terms in the literature can easily induce a future Big Rip singularity even if dark energy satisfies the null energy condition. In order to show the appearance of other types of singularities within our framework, we will consider a slightly modified version of the interaction term given by
\be
Q=\Gamma\rho_{DE}^n \ ,
\ee
with $n$ a dimensionless constant and $\Gamma$ a parameter with dimension $(\rm mass)^{5-4n}$  controlling the strength of the interaction term. Again, the interaction can be expressed in terms of the Hubble expansion rate as
\be
Q=\Gamma\left(\frac{3H^2+2\dot{H}}{\kappat^2}\right)^n  .
\ee
With this expression we can again obtain a differential equation for $H$ that will determine the cosmological evolution given by
\bea
2\ddot{H}+6(2+w_{DE})H\dot{H}+9(1+w_{DE})H^3\nonumber\\+\kappat^{2(1-n)}\Gamma\left(3H^2+2\dot{H}\right)^n=0 \ ,
\eea
where we have taken $\dot{w}_{DE}=0$ for simplicity again. It is interesting to notice that this system has de Sitter critical points (in addition to the Minkowski critical point) determined by
\bea 
H_{\rm dS}=\left[-\frac{9 (1+w_{DE})}{3^n \kappat^{1-n}\Gamma}\right]^\frac{1}{2n-3} \ ,
\eea
which exists for dark energy models with $w_{DE}\neq-1$. Since we are interested in obtaining additional future singularities (other than Big Rip) we will now look for solutions where $\vert\dot{H}\vert\gg H^2$. Furthermore, we seek for solutions driven by the interaction term so that we will also assume that $\vert H\vert\ll\vert \Gamma (\dot{H}/\kappat^2)^{n-1}\vert$ so that the above equation reduces to
\be
\ddot{H}+\mu\dot{H}^n\simeq0 \ ,
\label{Eq:asymp}
\ee
with
\be
\mu\equiv \Gamma\left(\frac{\kappat^2}{2}\right)^{1-n} \ .
\ee
Notice that the above equation is invariant under a constant shift of $H$ and this is important to keep $H$ finite. In fact, the above equation can be easily solved to give the following asymptotic solution of the original equation:
\be
H\simeq C(t_s-t)^p+H_s \ ,
\ee
with $t_s$ and $H_s$ integration constants and
\be
p=\frac{n-2}{n-1}, \quad\quad C=\frac{(1-n)^p}{n-2}\mu^{p-1} \ .
\ee
This shows that the considered interaction term can induce singularities where $H$ remains constant while its derivative diverges. For instance, if we take $n=3$ we find solutions of the form $H\simeq C\sqrt{t_s-t}+H_s$ which give $H(t_s)=H_s$ but $\dot{H}\to \infty$ as $t\to t_s$, i.e., a type II or Sudden Singularity. This behaviour will be general for values of $n$ leading to $0<p<1$. This type of solutions are explicitly shown in Fig. \ref{phasemapSudden} (upper panels), where we can indeed confirm the analytical asymptotic behaviour for the solutions. Furthermore, these interactions also allow to find solutions of a Big Freeze or type III singularities by imposing, for instance, $p=-1/2$, which is achieved for $n=5/3$. In that case $H$ diverges as $H\simeq C(t_s-t)^{-1/2}$ at the singularity (as well as its derivatives), but the scale factor approaches the singularity as $a\simeq a_s e^{2C\sqrt{t_s-t}}$ and, thus, it remains finite. This behaviour  will be typical for values of $n$ giving $-1<p<0$. Our analytical findings can be confirmed in Fig. \ref{phasemapSudden} (upper panels) from the phase map and explicit numerical solutions.

\section{Conclusions} \label{sec:V}

In this work we have revisited the so-called interacting dark energy models, where a coupling to dark matter is assumed, and we have established a general relation between such interactions and future cosmological singularities. We have considered the usual interaction terms at the level of the continuity equations so that the total energy is automatically conserved but a flow between both dark components exists. Within this scenario we have found that every future cosmological singularity taking place at a finite time can be directly mapped into a singularity of the interaction term,  which we have dubbed as $Q$-singularity. This means that the energy flow diverges at finite time, naturally inducing  one of the future singularities studied in the literature so far. Furthermore, our framework has allowed us to find a novel type of singularity characterized by a divergence in the time-derivative of the equation of state parameter of dark energy. Although this singularity is expected to not be relevant for the background evolution, it might signal the presence of divergences in the sound speed of the perturbations. We have exemplified these relations by considering parameterizations of the Hubble expansion rate and the dark energy equation of state.

As specific realizations of our general framework, we have also investigated the potential occurrence of the future singularities and their relation to $Q$-singularities when assuming some specific interacting terms given in terms of the energy density of dark energy. For usual interacting terms proportional to the energy density of dark energy, we have shown that the interaction with dark matter can induce a future Big Rip singularity even if the dark energy component does not violate the null energy condition, i.e., the interacting term can induce an effective phantom behavior for dark energy even if $w_{DE}>-1$. We have also considered interactions given in terms of an arbitrary power of the dark energy density and we have found that this interaction lead to other types of singularities such as Sudden and Big Freeze singularities. We have rigorously shown this by performing an analytical and a phase map analysis, whose results have also been confirmed by means of numerical solutions.

As a main result of our study we can conclude that interacting dark energy models provide a promising and very suitable framework to study cosmologies with future singularities, since they all can be accommodated in appropriate interaction terms, i.e., as $Q$-singularities.

\section*{Acknowledgments}

J.B.J. acknowledges the financial support of A*MIDEX project (No.~ANR-11-IDEX-0001-02), funded by the Investissements dAvenir French Government program, managed by the French National Research Agency (ANR);  MINECO  (Spain)  projects FIS2014-52837-P and Consolider-Ingenio MULTIDARK
CSD2009-00064. D.R.G. is funded by the Funda\c{c}\~ao para a Ci\^encia e a Tecnologia (FCT, Portugal) postdoctoral fellowship No.~SFRH/BPD/102958/2014. D.S.G. is funded by the Juan de la Cierva program (Spain) No.~IJCI-2014-21733 and FCT (Portugal) No.~SFRH/BPD/95939/2013. V.S. is funded by the Polish National Science Center Grant  DEC-2012/06/A/ST2/00395. The authors also acknowledge support from the FCT research grant UID/FIS/04434/2013. This article is based upon work from COST Action CA15117, supported by COST (European Cooperation in Science and Technology).

\end{document}